\documentclass[aps,pra,reprint,groupedaddress,showkeys]{revtex4-1}
\usepackage{amsmath,graphicx,verbatimbox,array,float}
\usepackage{indentfirst}
\usepackage{braket}
\usepackage{diagbox}
\usepackage{amssymb}
\usepackage{bbold}
\usepackage{colortbl}
\usepackage{multirow}
\usepackage{subcaption}
\newcommand{\figref}[1]{Fig.~\ref{#1}}
\newcommand{\tabref}[1]{Tab.~\ref{#1}}
\renewcommand{\eqref}[1]{Eq.~(\ref{#1})}
\newcommand{\tr}{{\mathrm{Tr}}}

\begin{document}
\title{Entanglement quantification enhanced by dark count correction}
\author{Artur Czerwinski}\email[]{aczerwin@umk.pl} 
\affiliation{Institute of Physics, Faculty of Physics, Astronomy and Informatics \\ Nicolaus Copernicus University in Torun, ul. Grudziadzka 5, 87--100 Torun, Poland}

\begin{abstract}
In this letter, we propose a method of dark count correction in quantum state tomography of entangled photon pairs. The framework is based on a linear model of dark counts, which is imposed on the measurement formalism. The method is tested on empirical data derived from an experiment on polarization-entangled photons. We demonstrate that the numerical approach to dark count correction guarantees more reliable state reconstruction compared with standard estimation. Most importantly, however, the conceptually simple theoretical approach proves to be more efficient at entanglement quantification than experimental techniques.
\end{abstract}
\maketitle

\section{Introduction}

The recent development of quantum-based technologies has made crucial the ability to characterize and validate quantum components. It can be done on various levels, including quantum state tomography (QST) \cite{Paris2004}, quantum process tomography \cite{Kliesch2019,Tran2021}, and detector tomography \cite{Lundeen2009,Consul2020}. For years, much attention has been paid to QST, which aims at experimentally determining a mathematical representation of a quantum state. Therefore, QST answers fundamental questions about the quantum world by providing all accessible information about a system. Such a procedure can be applied to characterize different physical systems. In the present work, we focus on reconstructing the polarization state of light \cite{James2001,Altepeter2005}.

Photons can be considered a versatile quantum resource since information can be encoded by exploiting various degrees of freedom such as spectral, spatial, temporal, and polarization. In addition, there are multiple protocols that take advantage of non-classical correlations that a pair of photons may feature \cite{Bell1964,Clauser1969}. Quantum entanglement \cite{Horodecki2009}, which is a prominent form of such correlations, has been applied to cryptography \cite{Ekert1991}, dense coding \cite{Bennett1992}, or teleportation \cite{Bennett1993}. In recent years, many practical implementations of quantum key distribution (QKD) protocols with polarization-entangled photon pairs have been proposed \cite{Jennewein2000,Poppe2004,Liu2010,Bedington2017}.

Before practical applications, a source of entangled photon pairs needs to be thoroughly investigated to evaluate the quality of entanglement. One way to achieve this involves QST, which provides the density matrix that represents the quantum state of a photon pair \cite{White1999}. Having the density matrix, one can compute an entanglement monotone, such as the concurrence \cite{Hill1997,Wootters1998}, to quantify the entanglement of the system. The approach based on complete state reconstruction can be an alternative to performing a CHSH-type Bell inequality test, cf. \cite{Nomerotski2020}. The main advantage of entanglement quantification via QST involves the ability to compute different characteristics of the state and conduct additional analysis.

In practical realizations of QST, noise and errors are inevitable and should be properly addressed. Numerical simulations can be helpful to assess the impact of selected types of errors on QST \cite{Czerwinski2022}. In experiments based on photon counting, one encounters, for example, the shot noise, dark counts, or uncertainty associated with detectors. In this work, we consider polarization measurements affected by dark counts.

In Sec. \ref{framework}, we introduce a framework for state estimation and entanglement quantification of photon pairs. The standard estimation technique is enhanced by a linear model that enables dark count correction. In Sec. \ref{resana}, we present the results obtained for polarization-entangled photon pairs. The findings confirm empirically that a theoretical method for dark count reduction can be more efficient at entanglement quantification than an experimental-based approach.

\section{Framework for state estimation and entanglement quantification}\label{framework}

\subsection{Methods}

Quantum tomography of states encoded on photons requires an estimator that can determine a density matrix that optimally fits the experimental data. In this context, we talk about measured and expected photon counts, denoted by $m_k$ and $e_k$, respectively, where the index $k$ refers to the index of the measurement operator. The measured photon counts are obtained from an experiment by adjusting the setup so that the number of photons that reach the detector can be interpreted as a result of quantum measurement.

In the present work, we deal with tomography of quantum states encoded in the polarization degree of freedom. For a single photon, the measurement scheme is based on six projectors $\{\mathcal{P}_i\}$ that correspond to horizontal, vertical, diagonal, anti-diagonal, left-circular, and right-circular. The projectors can be denoted as

\begin{equation}\label{sixM}
\begin{split}
&\mathcal{P}_1 = \ket{H}\!\bra{H}, \hspace{0.45cm} \mathcal{P}_2 = \ket{V}\!\bra{V}, \hspace{0.425cm} \mathcal{P}_3 = \ket{A}\!\bra{A},\\
&\mathcal{P}_4 = \ket{D}\!\bra{D}, \hspace{0.5cm}
\mathcal{P}_5 = \ket{L}\!\bra{L}, \hspace{0.5cm} \mathcal{P}_6 = \ket{R}\!\bra{R},
\end{split}
\end{equation}
where we assume that $\{\ket{H},\ket{V}\}$ form a standard basis in the two-dimensional Hilbert space. Qubit tomography with the six projectors \eqref{sixM} guarantees that the effects of statistical error are minimized \cite{Wootters1989}.

To reconstruct a quantum state of a two-qubit state, we implement product operators: $M_k := \mathcal{P}_{i} \otimes \mathcal{P}_{j}$ where, for simplicity, we use one index $k\equiv (i,j)$. Consequently, we obtain a set of $36$ two-qubit measurement operators $\{ M_k \}$. This measurement scheme can be considered overcomplete, but it is often applied in realistic situations. More kinds of measurement allow for a more reliable state estimation.

The set of measured photon counts $\{m_k\}$ comes from an experiment demonstrating that polarization-entangled photon pairs can be produced via the process of spontaneous parametric down-conversion (SPDC) by a photonic device called a Bragg reflection waveguide (BRW) \cite{Horn2013}. Implemented in gallium arsenide, this miniature chip-scale instrument can directly produce polarization-entangled photon pairs without additional path difference compensation, spectral filtering, or post-selection. Different measurements were performed on the photon pairs emerging from the source to demonstrate non-classical correlations in their polarization states. For example, the overcomplete set of $36$ polarization measurements was implemented to reconstruct the density matrix and quantify entanglement \cite{Horn2013}. In the present paper, we reuse the photon counts collected in that experiment to test numerical frameworks of state estimation.

The expected photon counts $e_k$ are modeled mathematically based on the knowledge we have about the measurements and other experimental factors. The simplest way to express the expected result of measurement is to follow the Born rule:
\begin{equation}\label{expected}
    e_k = \mathcal{N} \: \tr \left( M_k \rho  \right),
\end{equation}
where $\mathcal{N}$ denotes the total number of photon pairs emitted per measurement, and $\rho$ represents the unknown two-qubit state. Normally, we cannot precisely compute $\mathcal{N}$. However, we assume that each measurement is conducted with the same total number of photons. For example, in the experiment from Ref.~\cite{Horn2013}, each measurement had a duration of 2 minutes, which, combined with the constant power of the source, implies an equal number of photon pairs per measurement. Therefore, we treat $\mathcal{N}$ as an additional parameter that needs to be estimated in the framework. As for the density matrix, $\rho$, we have no \textit{a priori} information about its properties. Thus, we follow the most general representation of a two-qubit density matrix:
\begin{equation}\label{m5}
\rho = \frac{W^{\dagger} W}{\tr\: \left(W^{\dagger} W\right)},
\end{equation}
where $W$ denotes a left-triangular matrix:
\begin{equation}\label{m6}
W=\begin{pmatrix} w_1 & 0 & 0 &0 \\ w_5 + i\, w_6 & w_2 & 0 &0 \\  w_{11} + i \,w_{12} & w_7 + i\, w_8 & w_3 &0 \\ w_{15} + i\, w_{16} & w_{13} + i\, w_{14} & w_9 + i\, w_{10} & w_4 \end{pmatrix}.
\end{equation}
This parametrization, known as the Cholesky decomposition, guarantees that any density matrix computed from a tomographic framework is physical, i.e., it is Hermitian, positive semi-definite, of trace one, cf. \cite{James2001,Altepeter2005}.

The approach based on \eqref{expected} requires the ability to estimate the values of $16$ real parameters: $\mathcal{W} = \{w_1, w_2, \dots, w_{16}\}$ that completely characterize an unknown state $\rho$ as well as one additional parameter denoting the number of photon pairs in a beam, i.e., $\mathcal{N}$. However, \eqref{expected}, which is a standard formula for quantum measurement, does not involve any clue about possible unintended counts. Therefore, if we attempt to conform the measured counts (affected by dark counts) to the theoretical model that assumes noiseless measurements, we risk obtaining an inadequate density matrix. This implies that we need to reformulate \eqref{expected} to make it suitable for realistic state estimation.

We assume that the photonic signal that arrives at the detection systems consists of $\mathcal{N}$ photon pairs, but some of them are not the intended signal and can be described by the maximally mixed state. Thus the state that undergoes measurement can be given as:
\begin{equation}
    \tilde{\rho} = (1 - a) \,\rho + a\, \rho_{mix},
\end{equation}
where $\rho_{mix} = 1/4 \:\mathbb{I}_4$ and $0\leq a \leq 1$, which can be called the dark count rate. Also, we allow for a background noise, which is constant and independent of the source. Therefore, we include a second parameter that accounts for the average rate of registered counts without any incident light. Then, the expected photon count can be modeled as:
\begin{equation}\label{expected2}
\begin{aligned}
   \widetilde{e}_k  \,&=   \mathcal{N} (1 -a )\, \tr \left( M_k \rho  \right) + a \mathcal{N} \,\tr \left( M_k \rho_{mix}  \right)  + b = \\& 
=\mathcal{N} (1 -a) \,\tr \left( M_k \rho  \right) + a \frac{\mathcal{N}}{4} + b,
\end{aligned}
\end{equation}
where $b>0$ describes the false detection events. The formula for the expected photon counts \eqref{expected2} comprises a linear model of dark counts. By following \eqref{expected2}, one has to determine two additional parameters, $a$ and $b$, as compared to \eqref{expected}. However, we have at our disposal an overcomplete set of measured photon counts. It means that we can expect to be able to successfully implement the framework with extra parameters.

The set of $\mathcal{W}$ as well as any other parameters can be determined by minimizing selected estimators. We implement and compare the performance of three estimators: maximum likelihood (MLE) \cite{Ikuta2017}, $\chi^2$ \cite{Berkson1980,Jack2009,Bayraktar2016}, and least squares (LS) \cite{Acharya2019}, see \tabref{estymatory}.

\begin{table}[h]
\centering
\bgroup
\def\arraystretch{3.5}
	\begin{tabular}{|c|c|}
\hline
			\hspace{0.5cm}Estimator	\hspace{0.5cm} &  Formula \\ \hline
	  MLE  & $\mathcal{L}_{MLE} = \sum_{k} \left[\frac{(m_k -  e_k)^2}{ e_k }  + \ln e_k \right]$  \\ \hline

	 $\chi^2$ & $\mathcal{L}_{\chi^2} = \sum_{k} \frac{(m_k -  e_k)^2}{ e_k }$ \\
	 \hline
	LS & $\mathcal{L}_{LS} = \sum_{k} (m_k -  e_k)^2$ \\ \hline
	\end{tabular}
\egroup
	\caption{Estimators used to project the measurement data onto a parameter space.}
	\label{estymatory}
\end{table}

Solving the quantum state estimation problem relies on minimizing the formula for the respected estimator over all possible parameters. Each estimator from \tabref{estymatory} can be fed with either the standard expected counts \eqref{expected} or the formula that involves the linear dark count model \eqref{expected2}. For better differentiation, in the latter case, the corresponding estimators can be denoted by $\widetilde{\mathrm{MLE}}$, $\widetilde{\chi^2}$, and $\widetilde{\mathrm{LS}}$.
 
 The tomographic scheme is implemented in \textit{Wolfram Mathematia}, where we use Numerical Nonlinear Global Optimization algorithms for finding constrained global minima of the estimators. To be more specific, we utilize the \textit{NMinimize Function} with a stochastic function minimizer, which is called \textit{Differential Evolution}. In each run, the algorithm performs $1600$ iterations to find the global minimum. Additionally, to evaluate the uncertainty related to the numerical fitting, we perform the algorithm in $40$ independent trials since each time the method may converge to a different solution. Consequently, we obtain $40$ resulting density matrices, for which we can compute different measures. The procedure is repeated for every estimator and for each model of expected photon counts \eqref{expected} and (\ref{expected2}),
 
\subsection{Performance analysis}

We implement four figures of merit to evaluate the performance of the framework and compare the findings of the present analysis with the original results published in Ref.~\cite{Horn2013}.

Most of all, we use the concurrence the quantify entanglement of the reconstructed state $\rho$ \cite{Hill1997,Wootters1998}. To do so, we first compute the spin-flipped state, denoted by $\varrho$, which is defined as:
\begin{equation}\label{m11}
\varrho := \left(\sigma_y \otimes \sigma_y \right) \rho^* \left(\sigma_y \otimes \sigma_y \right),
\end{equation}
where $\rho^*$ represents the complex conjugate of $\rho$ and $\sigma_y$, by convention, denotes one of the Pauli matrices. Next, we construct the $R-$matrix:
\begin{equation}\label{m12}
R := \sqrt{\sqrt{\rho} \,\varrho\, \sqrt{\rho}},
\end{equation}
which leads to the definition of the concurrence, $C(\rho)$, which can be given by means of the eigenvalues of the $R-$matrix:
\begin{equation}\label{m13}
C(\rho) := \max \left\{0, \alpha_1 - \alpha_2 - \alpha_3 -\alpha_4   \right\},
\end{equation}
where $\alpha_1, \alpha_2, \alpha_3, \alpha_4$ are the eigenvalues of the $R-$matrix arranged in the decreasing order.

Next, we follow the standard definition of the purity to measure how much a state is mixed \cite{Nielsen2000}.

Furthermore, we employ the square root fidelity \cite{Jozsa1994,Uhlmann1976}:
\begin{equation}
    \mathcal{F} (\rho) := \tr \sqrt{\sqrt{\rho} \, \sigma  \, \sqrt{\rho}},
\end{equation}
which allows us to compute the overlap between the estimated density matrix $\rho$ and any given quantum state $\sigma$. Particularly, we consider two specific situations. First, we analyze the fidelity with one of the celebrated Bell states:
\begin{equation}
    \ket{\Phi^+} = \frac{1}{\sqrt{2}} \left(\ket{HV} + \ket{VH}  \right),
\end{equation}
which is the expected state produced by the source. Secondly, we compute the fidelity with the result presented in Ref.~\cite{Horn2013}, which was obtained by applying the Kalman filter and is denoted by $\rho_{K}$. The two fidelities are denoted by $\mathcal{F}_{B} (\rho)$ and $\mathcal{F}_{K} (\rho)$, respectively. In \tabref{figuresofmerit}, a brief summary of the figures of merit is presented.

\begin{table}[h]
\centering
\bgroup
\def\arraystretch{3.5}
	\begin{tabular}{|c|c|}
\hline
			\hspace{0.5cm}Figure \hspace{0.5cm} &  Formula \\ \hline
	 Concurrence  & $C(\rho) = \max \left\{0, \alpha_1 - \alpha_2 - \alpha_3 -\alpha_4   \right\}$ \\ \hline

	Purity & $\gamma(\rho) = \tr \rho^2$ \\
	 \hline
	Fidelity $1$ & $\mathcal{F}_{B} (\rho) = \bra{\Psi^+}\rho \ket{\Psi^+}$ \\ \hline
		Fidelity $2$ & $\mathcal{F}_{K} (\rho) = \tr \sqrt{\sqrt{\rho} \, \rho_{K}  \, \sqrt{\rho}} $ \\ \hline
	\end{tabular}
\egroup
	\caption{Figures of merit used to evaluate the performance of the framework.}
	\label{figuresofmerit}
\end{table}

Each figure of merit is averaged over the set of $40$ density matrices obtained in a given setting. Then, the mean value with the sample standard deviation (SD) is considered as an indicator of the performance.

\section{Results and analysis}\label{resana}

The figures of merit are gathered in \tabref{results}. First, for comparison, we recall the results presented in Ref.~\cite{Horn2013}. Those outcomes were generated by applying the Kalman filtering, cf. \cite{Audenaert2009}. Additionally, in Ref.~\cite{Horn2013}, background count rates were collected by changing the electronic delay in the idler arm to facilitate the observation of random background events. This experimental approach allowed the authors to subtract the dark counts, which resulted in the concurrence $0.52$. In this section, we can compare the efficiency of the experimental treatment of dark counts with the numerical model for noise correction.

\begin{widetext}
\begin{table}[h!]
\centering
\bgroup
\def\arraystretch{3.0}
	\begin{tabular}{|c|c|c|c|c|c|c|c|}
\hline
			\backslashbox[22mm]{\color{black}Figure}{\color{black}Method} &  Kalman Filter \cite{Horn2013} & MLE & $\widetilde{\mathrm{MLE}}$ & $\chi^2$ & $\widetilde{\chi^2}$ &  LS & $\widetilde{\mathrm{LS}}$ \\ \hline
	  $\;C(\rho)\;$  & $\; 0.52 \;$  & \hspace{0.175cm}$0.372489(1)$\hspace{0.175cm} & \hspace{0.175cm}$0.578255(1)$\hspace{0.175cm} & \hspace{0.175cm}$0.371755(1)$\hspace{0.175cm} & \hspace{0.175cm}$0.578255(1)$ \hspace{0.175cm}& \hspace{0.175cm}$0.453378(1)$ \hspace{0.175cm}& \hspace{0.175cm} $0.537686(1)$ \hspace{0.175cm}\\ \hline

	 $\;\gamma(\rho)\;$ & $ 0.64$ & $0.552242(1)$ & $0.646282(1)$ & $0.552108(1)$ & $0.646282(1)$ & $0.570081(1)$& $0.606781(1)$\\
	 \hline
	$\;\mathcal{F}_{B} (\rho)\;$ & $0.83$ & $0.798494(1)$ & $0.838746(1)$ & $0.798426(1)$ & $0.838746(1)$& $0.806616(1)$& $0.821085(1)$\\ \hline

		 $\;\mathcal{F}_{K} (\rho)\;$ & -- & $0.98615(1)$ & $0.999138(1)$ & $0.986072(1)$ & $0.999138(1)$  & $0.993387(1)$& $0.996986(1)$\\ \hline
	\end{tabular}
\egroup
	\caption{Figures of merit for six methods of state estimation. Each result is accompanied by SD.}
	\label{results}
\end{table}
\end{widetext}

First, we see that the estimators MLE, $\chi^2$, and LS, which were filled with the standard formula for the expected counts \eqref{expected}, delivered significantly inferior results to the Kalman Filter. This was an anticipated outcome since the formula \eqref{expected} does not convey any information about possible photon noise. However, it appears intriguing that out of the three estimators, we get the best figures of merit with the method of least squares, which uses the simplest estimator. By comparison, we see that the values corresponding to MLE and $\chi^2$ are close to each other, but for LS, we have significantly better performance. In particular, if we consider the concurrence, we notice that LS has led to states that feature more entanglement. Nonetheless, in the standard approach, MLE, $\chi^2$, and LS fall behind the Kalman Filter, which took into account experimental dark counts.

The key point of this analysis relates to the results that were obtained by implementing the extended formula for the expected counts \eqref{expected2}. The findings of this scenario are denoted in \tabref{results} by $\widetilde{\mathrm{MLE}}$, $\widetilde{\chi^2}$, and $\widetilde{\mathrm{LS}}$, for the corresponding estimators. First, one can notice that $\widetilde{\mathrm{MLE}}$ and $\widetilde{\chi^2}$ provided exactly the same figures of merit. This means that both estimators are equally efficient. On the other hand, $\widetilde{\mathrm{LS}}$ turned out the be the least productive.

It is worth highlighting that the three estimators: $\widetilde{\mathrm{MLE}}$, $\widetilde{\chi^2}$, and $\widetilde{\mathrm{LS}}$, which involve the linear model of dark counts, outperformed the Kalman Filter when it comes to entanglement quantification. The difference is particularly evident for $\widetilde{\mathrm{MLE}}$ and $\widetilde{\chi^2}$, when the relative rise of concurrence is $11,2\,\%$. This outcome proves that the numerical model of dark count correction can be more efficient at entanglement quantification than experimental measurements of dark counts. Thanks to the overcomplete set of measurements, we are capable of estimating not only the density matrix, but also the parameters that describe dark counts according to the linear model.

\begin{figure}[h]
	\centering
  \begin{subfigure}{1\columnwidth}
\caption{the real part of $\tilde{\rho}$}
       \centering
         \includegraphics[width=0.95\columnwidth]{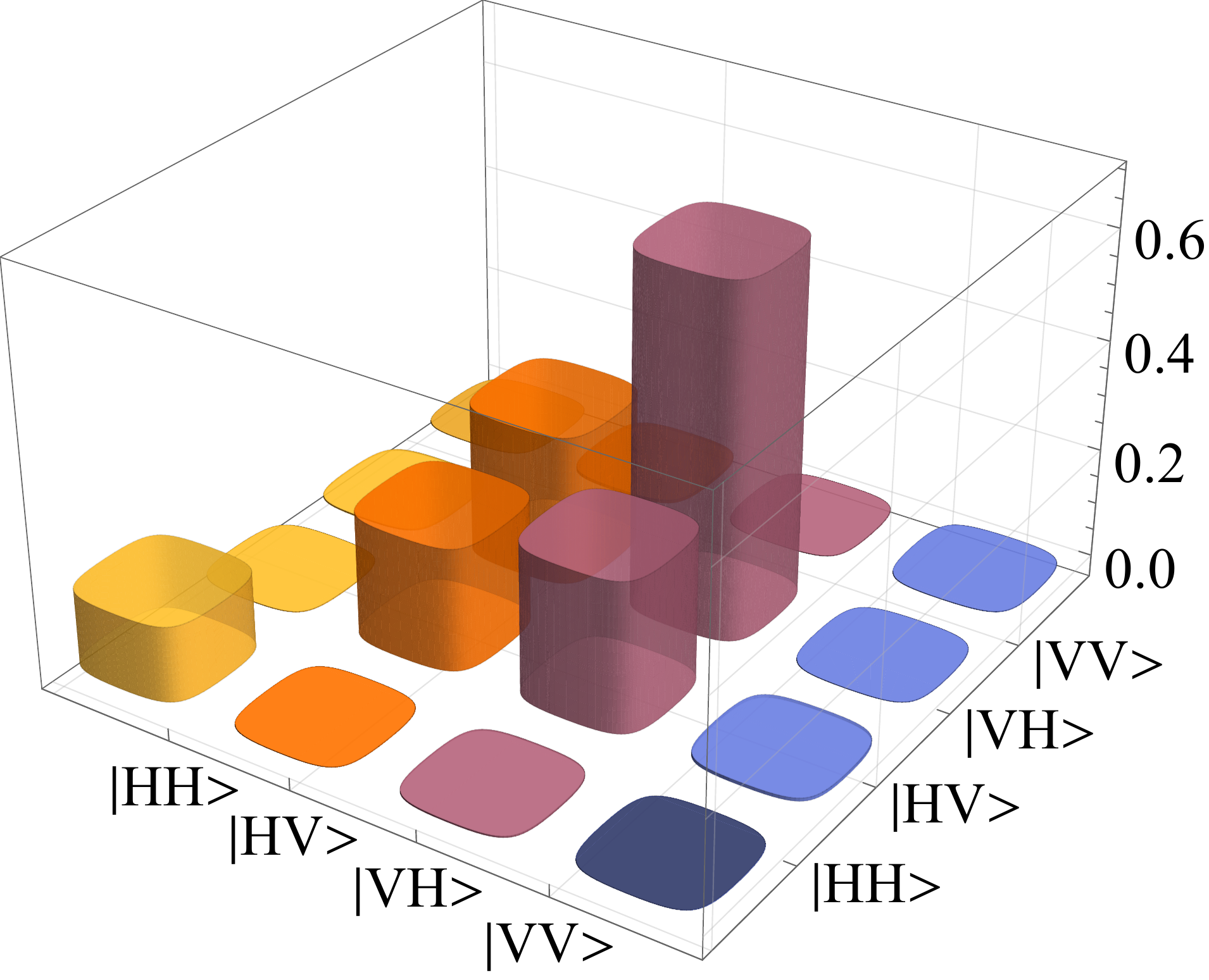}
\label{rerho}
     \end{subfigure}
     \hfill
     \begin{subfigure}{1\columnwidth}
\caption{the imaginary part of $\tilde{\rho}$}
       \centering
         \includegraphics[width=0.95\columnwidth]{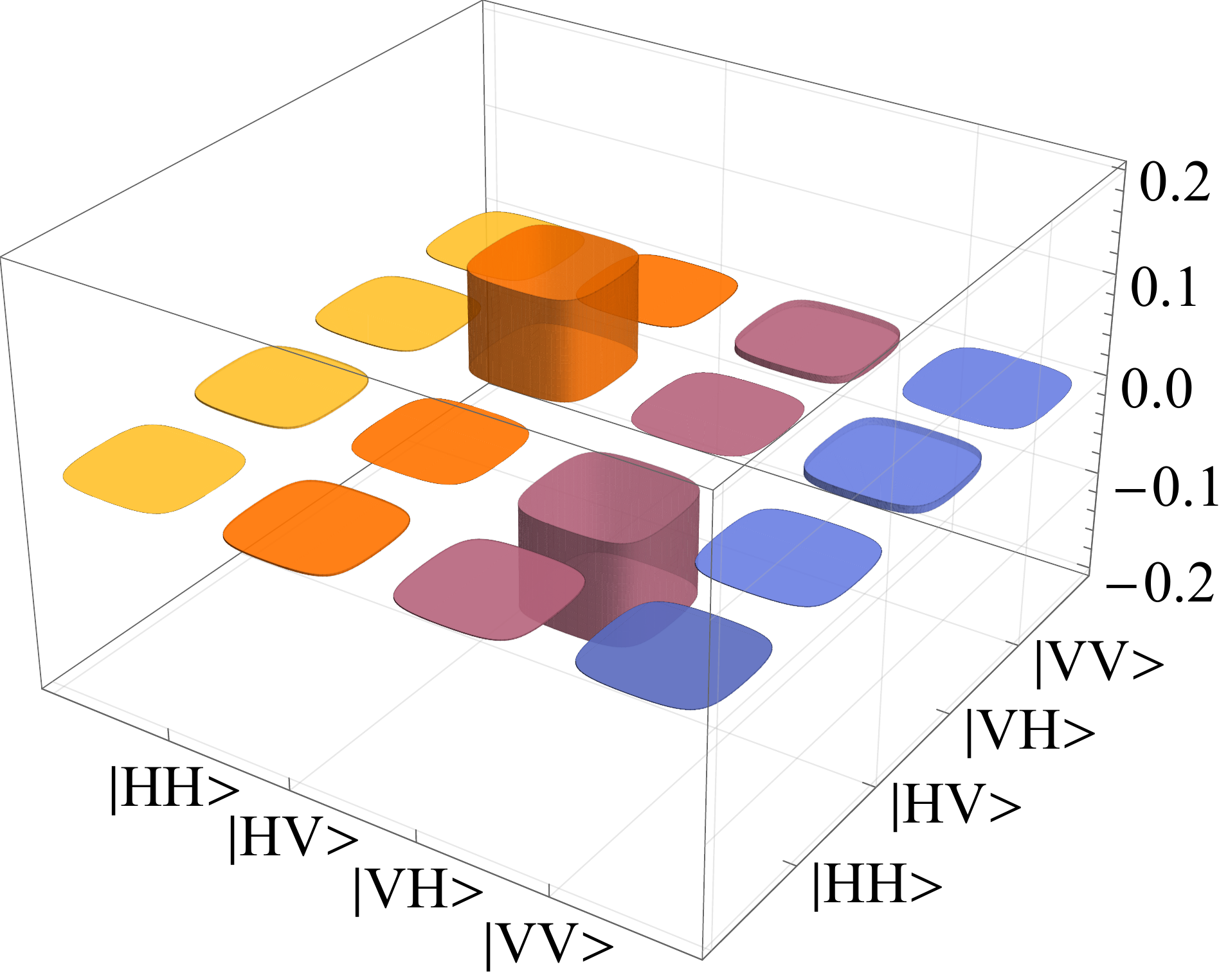}
\label{imrho}
     \end{subfigure}
	\caption{Graphical visualization of the density matrix provided by $\widetilde{\mathrm{MLE}}$.}
	\label{bestmatrix}
\end{figure}

\begin{figure}[h]
\centering
\includegraphics[width=0.95\columnwidth]{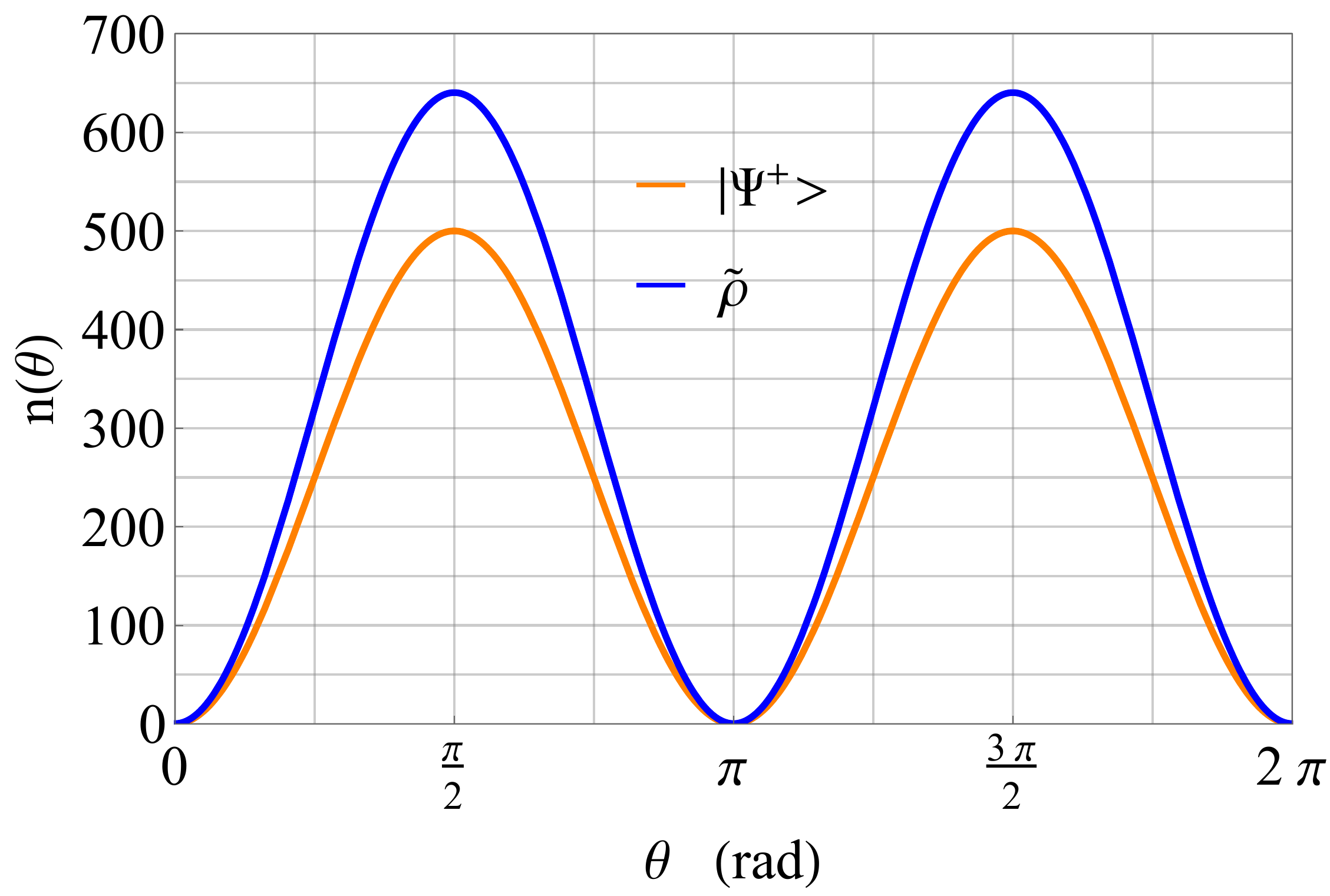}
\caption{Polarization entanglement analysis \eqref{analysis}. The number of photon pairs is fixed: $\mathcal{N} = 1\,000$.}
\label{coincidences}
\end{figure}

Furthermore, the optimal results of state estimation can be visualized graphically, which is shown in \figref{bestmatrix}. The matrix presented in the plots was computed by averaging over the states provided by $\widetilde{\mathrm{MLE}}$, i.e., $\tilde{\rho} = 1/40 \,\sum_{j=1}^{40} \rho_j$, where $\{\rho_j\}$ is the set of estimates brought by $\widetilde{\mathrm{MLE}}$. The visualization allows us to observe the differences between $\tilde{\rho}$ and the target state $\ket{\Psi^+}$.

Finally, the last step of the study involves polarization entanglement analysis for the state $\tilde{\rho}$, compared to $\ket{\Psi^+}$. Such an analysis can be realized, assuming
that the polarization analyzer in one arm is fixed at the vertical (V) configuration, whereas the idler analyzer rotates to perform all linear polarization measurements. We investigate the function of measurement that represents the average number of detected coincidences:
\begin{equation}\label{analysis}
    n (\theta) = \mathcal{N} \, \tr \left( \ket{V}\!\bra{V}\otimes \ket{\theta}\!\bra{\theta} \,\rho  \right),
\end{equation}
where $\ket{\theta} = (\sin \theta \; \cos \theta)^T$ and $0\leq \theta < 2 \pi$. In \figref{coincidences}, we present the plots of $n (\theta)$, where $\rho$ was substituted with either $\tilde{\rho}$ or $\ket{\Psi^+}\!\bra{\Psi^+}$. The plots display non-classical correlations of $\tilde{\rho}$ but also facilitate the comparison between the estimated state and the target one.

First, we notice that the roots of both plots overlap, which is related to the fact that for $\ket{\Psi^+}$, there is zero probability of finding both photons in the vertical polarization. This property has been well retrieved by the framework, which can also be seen in \figref{bestmatrix}. However, the plots diverge as the function $n (\theta)$ approaches its maximum value. For $\ket{\Psi^+}$, $n_{max} = n (\pi/2) = 500$, which stems from the probability amplitude in the state vector. However, for the estimated state $\tilde{\rho}$, the probability of the event that one photon is found in the vertical polarization and the other in the horizontal is equal approx. $0.64$. Consequently, we observe a higher number of coincidences in the plot \figref{coincidences}. In spite of the inaccuracies, the analysis proves non-classical behavior of the state obtained via the framework.

\section{Conclusions and outlook}

In this paper, we have demonstrated that a numerical method of dark count correction can improve entanglement quantification compared with experimental-based techniques. The method employs a linear model of photon noise. The parameters that characterize the noise function are estimated along with the density matrix of the state in question. The framework was tested empirically with polarization-entangled photon pairs and proved to increase the concurrence of the estimated state by $11,2\,\%$ in comparison to the experimental reduction of dark counts.

The findings of the paper imply that if we have an overcomplete set of measurements at our disposal, we can determine not only the density matrix of the system but also estimate additional parameters that characterize the physical situation. This demonstrates that an overcomplete scheme can be more advantageous than a minimal set of measurement operators.

\section*{Acknowledgments}

The research was supported by the National Science Centre in Poland, grant No. 2020/39/I/ST2/02922.\\
The author acknowledges Dr. Piotr Kolenderski, who shared the data from the experiment \cite{Horn2013} and agreed to reuse it for present replications.

\end{document}